\documentclass[a4paper,conference]{IEEEtran} 
\ifCLASSINFOpdf
\else
\fi
\usepackage[cmex10]{amsmath}
\usepackage{amssymb,verbatim}
\usepackage{graphicx}

\begin{document}
%
\title{%
Application of quantum Pinsker inequality to quantum communications
}

\author{
\IEEEauthorblockN{Osamu Hirota\\}
\IEEEauthorblockA{Research and Development Initiative, Chuo University, \\
1-13-27, Kasuga, Bunkyou-ku, Tokyo 112-8551, Japan\\
Quantum ICT Research Institute, Tamagawa University\\
6-1-1, Tamagawa-gakuen, Machida, Tokyo 194-8610, Japan\\
{\footnotesize\tt E-mail: hirota@lab.tamagawa.ac.jp} \vspace*{-2.64ex}}
}

\maketitle

\begin{abstract}
Back in the 1960s, based on Wiener's thought, Shikao Ikehara 
(first student of N.Wiener) encouraged the progress of 
Hisaharu Umegaki's research from 
a pure mathematical aspect in order to further develop 
the research on mathematical methods of quantum information 
at Tokyo Institute of Technology.  Then, in the 1970s, 
based on the results accomplished by Umegaki Group, Ikehara 
instructed the author to develop and spread quantum information science 
as the global information science.  While Umegaki Group's results 
have been evaluated as major achievements in pure mathematics, 
their contributions to current quantum information science have not been fully discussed.  
This paper gvies a survey of my talk in the memorial seminar on Ikehara, 
in which  Ikehara and Umegaki Group's contributions to design theory 
of quantum communication have been introduced with specific examples 
such as quantum relative entropy and quantum Pinsker inequality.

\end{abstract}

%
\IEEEpeerreviewmaketitle
\section{Introduction}
In the real world, we have no performance evaluation measures for 
communication system with operational meanings of the global information 
transmission and processing other than various signal detection 
criteria established by Wiener, and Shannon entropy by Shannon.  
The same is true even where the physical system for implementation 
is generalized into quantum system or relativistic  
system, which means that modern communication theory is the most successful 
field among other scientific theories in human history.

Quantum information science originates from quantum communication 
theory, based on which, quantum communication such as quantum key 
distribution and quantum symmetric key cipher has been developed.  
These are formulated based on and as quantum versions of statistical 
signal detection theory and Shannon's information transmission theory.  
The former has a beautiful form as a design theory for detection 
and estimation techniques of signals transmitted in a quantum state 
established by Helstrom [1], Holevo [2] and Yuen [3].  The theory of 
Shannon information transmitted in a quantum state was started by 
Stratonovich, Holevo, et al and studied by Yuen [4] and Hirota [5] 
from the viewpoint of quantum state control as well as by Jozsa 
Group [6] and Masashi Ban of Tamagawa University Group [7] 
in the context of accessible information.

On the other hand, in mathematics, theories may be developed without 
regard for the operability of information handled by humans or with 
a focus on physical phenomena of a specific device.  Shannon's entropy 
 clearly defines very common signals processed in human 
social activities and it is applied to the communication system with 
the operational meanings for the relevant information processing.  
Quantum information theory as mathematics can be modeled on it.  
However, unlike Shannon information, quantum entropy lacks 
versatility regard for operational meaning on information transmission.  
It is defined in consideration of mathematical form 
or application to physics.  Therefore, no new information scientific  
technology is expected from simple generalization of the concepts 
of Wiener and Shannon.  Only when its 
contribution to Wiener-Shannon system is proved, the mathematics is 
deemed to have contributed to the global information science. 

It was Holevo who discovered a liaison between quantum entropy and 
Shannon information.  In the study of accessible information (
an application of Shannon's mutual information) to quantum 
system, its upper bound is now called Holevo bound, which is given 
in the form of quantum entropy, and its measure is called Holevo information.  
Holevo derived the upper bound in flow of the result of 
Stratonovich on $N$th extended system in relation to Shannon mutual 
information in quantum system (see Reference 8).  
Prior to that, Tamagawa University Group had published a paper 
providing specific examples of maximization of accessible information 
and super-additivity in  $N$th extended systems.  
Jozsa Group proved that the maximum amount of accessible information 
reaches Holevo information in the limit of pure quantum state 
system without external noise.  Jozsa presented the results 
at the 3rd International Conference on Quantum Communication, 
Measurement (QCM 1996).  While Holevo had shifted 
to research on quantum stochastic process, I recommended him that 
he should generalize Jozsa's works.  In a very short while, 
Holevo and Tamagawa University Group discussed the proof method 
in the general model of external noise system with Kitaev, and only after 
a week later, Holevo showed us the proposed proof on a white board 
at Tamagawa University.  We conveyed the results to Jozsa and Holevo 
submitted  his proof to IEEE's Transaction on Information Theory.  
Jozsa Group also started their consideration and the proof was completed by 
Shumacher-Westmoreland and submitted to the Physical Review.  
Although published at different times, these are now called 
Holevo-Shumacher-Westmoreland theorem as the formula of discrete 
channel capacity of classical-quantum composite system [9, 10].  

Additionally, Holevo and Tamagawa University Group gave a guide on 
formula for Gaussian channel capacity and for reliability function 
in 1996$\sim$2000 [11, 12].  As a result, Holevo information, 
which is expressed in the form of quantum entropy and serves 
as a parameter of the extreme point of Shannon system, has greatly 
contributed to the real world.

On the other hand, Umegaki and his group  
developed as mathematics in a quantum entropy form without 
considering the operational meaning like Shannon information.  
This paper will clarify Umegaki's contribution on the application of 
quantum information theory to real world issue.

\section{Progress in quantum entropy}
\subsection{Approach of  Umegaki and Holevo}
Let us here denote the most fundamental formula in Shannon's communication theory 
in the following.
\begin{eqnarray}
H(X)&=&-\sum_x P(x) \log P(x) \\
H(X|Y)&=&- \sum_y \sum_x P(y)P(x|y)\log P(x|y) \\
I(X,Y)&=& H(X)-H(X|Y)=H(Y)-H(Y|X) \nonumber \\
&=&H(X)+H(Y)-H(X,Y)
\end{eqnarray}
Shannon's information $H(X)$ and mutual information $I(X,Y)$ 
have a clear operational meaning according to the coding theorem. 
In physics, mutual information is sometimes regarded as a measure of 
classical correlation between two statistical systems. 
Meanwhile, entropy was introduced in 1951 by Kullback in statistics [13]. 
In that case, there is a strong intention to express the distance between 
probability distributions representing the characteristics of two 
statistical systems. Therefore, they start based on the following relative entropy:
\begin{equation}
D_c(P(x)||Q(x)) = \sum_x P(x)\log \frac{P(x)}{Q(x)}
\end{equation}

One can generalize the above into the composite system. If we denote 
the joint probability on the composite system as follows:
$P(x,y), Q(x,y)=P(x)P(y)$
we have
\begin{eqnarray}
D_c(P(x,y)||Q(x,y))&=&\sum_x \sum_y P(x,y) \log \frac{P(x,y)}{P(x)P(y)} 
\nonumber \\
&=&I(X,Y)
\end{eqnarray}
Thus, from a mathematical definition point of view, the relative entropy 
looks like $``General" $. 

On the other hand, von Neumann defined entropy for quantum systems 
in response to the development of quantum statistical mechanics. 
The entropy of a quantum system with the quantum density operator: \\
$ \rho_{X_q} \in {\cal D} (H_S) $ is
\begin{equation}
S(X_q)\equiv S(\rho_{X_q}) =-Tr\{\rho_{X_q}\log\rho_{X_q}\}
\end{equation}

This is called von Neumann entropy. Research on a quantum version of 
relative entropy, which is regarded as a mathematical generalization of 
Shannon's entropy theory, was begun for the first time in the world 
by Hisaharu Umegaki at Tokyo Institute of Technology. 
This was driven by the motivation of Shikao Ikehara to recommend the 
succession of Wiener thought [14,15].
Umegaki, for the first time in the world, defined the following quantum 
relative entropy on the von Neumann algebra in 1962 and formulated its 
various features [16-18].\\

${\bf {Definition-1(Umegaki)}}$\\
\begin{eqnarray}
&&D_q(\rho ||\sigma )=Tr \{\rho [\log \rho -\log \sigma ]\} \\
&& supp(\rho)  \subseteq supp(\sigma )
\end{eqnarray}

On the other hand, Helstrom and Holevo inherited the idea of Wiener- Shannon, 
and they defined that, in a communication system using quantum phenomena, 
the sender prepares a set of quantum density operators: \\
$ {\cal \epsilon} = \{p(x), \rho^x_{Y_q} \} $. 
That is, a message $x$ is mapped to a quantum density operator 
which corresponds to a concrete signal. 
The quantum density operator of the set is described as follows:
\begin{equation}
\rho_{Y_q}=\sum_x p(x) \rho^x_{Y_q}
\end{equation}
The receiving system performs a quantum measurement on the quantum system, 
and becomes a model for determining the classical parameter $\{x \}$ 
as a message. This is a problem of Accessible information 
(Shannon mutual information of classical-quantum composite systems).

\begin{equation}
I_{acc}=\max_{\Pi} I(X_c,Y_q)
\end{equation}
where $\Pi$ is detection operator or positive operator valued measure (POVM).
That is, to preserve Shannon's view of the world, Holevo considers the set whose 
signal element is the classical parameter of the quantum density 
operator, $ {\cal \epsilon} = \{p(x),\rho^x_{Y_q} \}$.
And the quantum entropy of the classical-quantum composite system 
was defined in 1973 as follows.\\

${\bf {Definition-2(Holevo)}}$
\begin{equation}
\chi({\cal \epsilon })=S(\rho_{Y_q})-\sum_x p(x)S(\rho^x_{Y_q})
\end{equation}
This is called the Holevo information. As stated in the introduction, 
the Holevo -Shumacher-Westmoreland theorem guarantees that the limit 
of Shannon information transmitted in a quantum system is the maximum 
value of Holevo information. This fact shows that quantum entropy 
theory contributes to the Shannon theory. 

As mentioned above, Umegaki developed with a foundation 
of mathematical statistics, and Holevo developed quantum communication theory 
while faithfully inheriting Shannon's world.

\subsection{More progress}
Although quantum entropy can be developed by mathematical formalism, 
there is no guarantee that it will have the operational meaning 
applicable in the real world like Shannon theory.  Wiener criticized 
simple mathematical generalization, claiming that unless a research 
directly transforms the technical system in the real world through 
mathematical generalization, it is not an authentic mathematical study.  
This is called Wiener's criteria for mathematical generalization research.  
This paper discusses whether quantum entropy theory satisfies 
Wiener's criteria.  

Based on Umegaki's quantum relative entropy, 
we can formally replace Shannon's formula with its quantum version.  
It was carried out faithfully under the development 
of quantum entropy theory as mathematics by collaborating with both 
Japanese and foreign researchers such as Accardi, Belavkin,Ohya and Petz [19].

Let us construct the Shannon's mutual information by quantum entropy form.
When the quantum density operator is 
$\rho_{X_qY_q} \in {\cal D}(H_X \otimes H_Y)$, we have 
\begin{equation}
S(X_q,Y_q)= -Tr\{\rho_{X_qY_q} \log \rho_{X_qY_q} \}
\end{equation}
Therefore, the quantum mutual information can be defined as follows.
\begin{equation}
I_{q}(X_q,Y_q)=S(X_q) +S(Y_q) -S(X_q,Y_q)
\end{equation}
However, since the quantum entropy is not information in the meaning 
of a message, the above expression does not have an operational 
meaning of a general communication system. On the other hand, 
mathematically, the Holevo information can be expressed 
in this context. Let us assume that the quantum density operator as a set of 
classical-quantum composite systems is given by
\begin{equation}
\rho_{X_qY_q}=\sum_x p(x)|x><x|_{X_q} \otimes \rho^x_{Y_q}
\end{equation}
So we have 
\begin{equation}
S(X_c,Y_q)=H(X_c) + \sum_x p(x) S(Y_q|x)
\end{equation}
From the above, also we have
\begin{equation}
\chi({\cal \epsilon })=I_{cq}(X_c,Y_q)
\end{equation}

In this way, the quantum mutual information contains formally the Holevo 
information, but the operational meaning is 
completely different, and only the Holevo information has 
significance for the Shannon system that has a great impact on 
the real world. On the other hand, 
from a mathematical point of view, the quantum mutual information 
can be expressed in terms of quantum relative entropy. That is,
\begin{equation}
I_{q}(X_q,Y_q)= D_q(\rho_{X_qY_q}||\rho_{X_q} \otimes \rho_{Y_q})
\end{equation}

Furthermore, the Holevo information is also described by 
quantum relative entropy as follows:
\begin{equation}
\chi({\cal \epsilon })=\sum_x p(x)D_q(\rho^x_{Y_q}||\rho_{Y_q})
\end{equation}

Thus, from the mathematical point of view, Umegaki's quantum relative entropy 
is the most fundamental notion. Based on this, quantum entropy theory 
has advanced rapidly in the 21st century as a mathematical study applying quantum 
statistical physics [20]. In the next section, we focus on applicability of the abstract mathematics to real communication sysytems based on Pinsker 
inequalities that give linkages to statistics and signal detection theory.

\section{Quantum Pinsker Inequality}
Relative entropy is essentially the distance between two probability 
distributions in statistics. Thus,it is natural to consider the relationship 
with various mathematical distances. In general, the distance between 
two probability distributions is called a statistical distance or 
Kolmogorov distance, and is defined as follows.
\begin{equation}
||P(x)-Q(x)||_c=\sum_x |P(x)-Q(x)|
\end{equation}

Such a concept of distance is often discussed in the language of 
distinguishability, and it is a source of great misunderstanding 
when one applies such mathematics to another problem. 
Here, it is discussed as a distance. 
The most important inequality in distance relations in statistics is 
the following Pinsker inequality shown by Pinsker in 1964. \\

${\bf {Theorem-1(Pinsker)}}$
\begin{equation}
D_c(P(x)||Q(x)) \geq \frac{1}{2\ln 2} ||P(x)-Q(x)||_c^2
\end{equation}

By utilizing this, generalization to the mutual information  
for the composite system becomes possible as follows:\\

${\bf {Theorem-2}}$
\begin{eqnarray}
&& I(X,Y) \geq \frac{2}{\ln 2} \Delta_c^2 \\
&& \Delta_c =\frac{1}{2}||P(x,y)- P(x)P(y)||_c
\end{eqnarray}

In quantum systems, the quantum density operator corresponds to the 
probability distribution in classical statistics.  
So the basic distance is ``trace distance" defined as follows:
\begin{equation}
\Delta_q =Tr\{\Pi^{opt}(\rho-\sigma)\}=\frac{1}{2}||\rho-\sigma ||_q
\end{equation}
where $ \Pi^{opt} $ is detection operator or a positive operator 
valued measure (POVM). 
The relationship between relative entropy and statistical distance 
shifts to the relationship between quantum relative entropy 
and trace distance. It was expressed  by the cooperation 
among Hiai, Ohya, and Tsukada as follows [21].\\

${\bf {Theorem-3(Quantum Pinsker Inequality)}}$
\begin{equation}
D_q(\rho||\sigma) \geq \frac{1}{2\ln 2} ||\rho-\sigma ||_q^2
\end{equation}
This can be further generalized to a quantum composite system. 
Let us assume quantum density operators in composite system as follows:
$\rho_{X_qY_q} \in {\cal D}(H_X\otimes H_Y)$,
$\Delta_q=1/2||\rho_{X_qY_q} - \rho_{X_q}\otimes \rho_{Y_q}||_q$
then we have the relation between quantum mutual information
$ I_q(X_q,Y_q)$ and trace distance.\\

${\bf {Theorem-4}}$
\begin{equation}
I_q(X_q,Y_q) \geq \frac{1\cdot }{\ln 2} 
||\rho_{X_qY_q} - \rho_{X_q}\otimes \rho_{Y_q}||_q^2=\frac{2}{\ln 2} \Delta_q^2
\end{equation}
\\
At this stage, quantum entropy theory does not play an important role 
in the Wiener-Shannon systems, and does not contribute as 
a design theory for real communication technologies. But in the next section, 
we will show remarkable result for applications to real world issue.

\section{Upper bound theory of guessing probability in QKD}
In the quantum entropy theory, only the Holevo information 
contributes to the Wiener and Shannon systems related to information 
communication systems, and it opened up the real world of optical quantum 
communication systems. On the other hand, in the context of Umegaki's research, the Holevo information can be formally expressed as 
a special example of quantum mutual information from Eq(16).
Here let us denote the trace distance as follows:
\begin{eqnarray}
\Delta_q&=&\frac{1}{2}||\sum_x p(x)|x><x|_{X_q} \otimes \rho^x_{Y_q} \nonumber \\
&&- \sum_x p(x)|x><x|_{X_q} \otimes \rho_{Y_q}||_q
\end{eqnarray}
Here we can show the following important theorem [22] based on 
Eqs(17, 18, 25, 26):\\

${\bf {Theorem-5}}$\\
The trace distance is bounded by Holevo Information as follows:
\begin{equation}
\chi({\cal \epsilon}) \geq \frac{2}{\ln 2} \Delta_q^2
\end{equation}

Even at this stage, the trace distance of the two quantum density 
operators in the above show only the characteristics of the quantum system, 
and the relationship with the evaluation of the technical operation 
in the Wiener-Shannon system is not clear. That is, the contribution 
to the real system is not visible.
In order to show that the theory of quantum entropy contributes to 
the Wiener-Shannon's communication theory, 
it is necessary to show that the trace distance defined before 
observation contributes directly to traditional 
performance evaluation measures in Wiener-Shannon system.

Before entering the main topic, we discuss with regard to the trace 
distance in the theory of quantum key distribution (QKD), 
because there is a theory that is misunderstood. It is supposed in 
QKD theory that there are quantum density operators formed by real 
protocols and quantum density operator formed by ideal protocols. 
They introduced Helstrom's quantum signal detection theory as a model 
to discriminate between these two quantum density operators and 
show the following average error probability or detection probability 
from the Helstrom formula.
\begin{eqnarray}
P_e&=&\frac{1}{2}[1-\Delta_q(\rho^R_{AE}, \rho^I_{AE})] \\
P_d&=&\frac{1}{2}[1+\Delta_q(\rho^R_{AE}, \rho^I_{AE})] 
\end{eqnarray}
where $ \Delta_q =\frac{1}{2} || \rho^R_{AE}-\rho^I_{AE} ||_q $ 
is the trace distance. 
In the first place, there is no physical communication system that transmits 
and receives real and ideal quantum density operators, 
so this model cannot be a tool for discussing the security of 
QKD. In addition, $\Delta_q $ is a parameter and cannot have a meaning as probability by itself. So the trace distance does not  contribute to security  
evaluation at this stage.

However, in 2009, Yuen provided a significant inequality.
That is, when the attacker accesses the number of signals 
$ M = 2^{| K |} $ with key sequence length $ | K | $ for 
real protocols in the context of the security of QKD,
for the statistical distance $ \Delta_c $ of 
the probability distributions after quantum measurement, 
the upper bound of the guessing probability is given as follows [23,24,25].\\

${\bf{Theorem-6}(Yuen)}$
\begin{eqnarray}
&&P(K)_{guess}\le \frac{1}{M} +\Delta_c \\
&&\Delta_c =\frac{1}{2}||P(x,y)- P(x)P(y)||_c \nonumber
\end{eqnarray}
\\
The author's group were able to conclude from the discussion with Yuen 
that the above relationship could be applied to the level 
of trace distance before making specific observations. 
The final expression is as follows [22,26].\\
\\

${\bf{Theorem-7 }}$\\
Let the trace distance of the quantum density operators between  
an actual protocol and the ideal one be:
\begin{eqnarray}
& &\Delta_q=\max_{\Lambda }Tr\Lambda (\sum_k p(k)\rho^k_{KE}-\rho_{K}
\otimes \rho_{E})\\
& &k \in {\cal M}, \quad \Lambda :POVM \nonumber
\end{eqnarray}
Then the average guessing probability for real QKD signals is 
\begin{equation}
\frac{1}{M} \le P(K)_{guess} \le \frac{1}{M} +\Delta_q
\end{equation}
where $\chi({\cal \epsilon}) \geq \frac{2}{\ln 2} \Delta_q^2$ from the theorem 5.
\\

The above equations can be obtained from the relationship 
between the statistical distance and the trace distance, but 
another direct proof is shown in the appendix for the convenience of the readers. 

From the results of Theorem 5 and Theorem 7, theory of Umegaki and his group contributes to the design theory of 
actual communication technologies in a sense different from 
the Holevo-Shumacher-Westmoreland theorem via the Holevo information. 

\section{Conclusion}
Quantum communication theory based on the basic concept of Wiener 
and Shannon has already contributed to the real optical communication 
system in a concrete manner [27, 28, 29].  I believe that this success 
originates from Ikehara's human resource development activities 
for mathematical basic research of quantum information, which were 
passed down to the later generations as part of Umegaki's extensive 
mathematical research.  I also believe that philosophies of Ikehara 
and Umegaki greatly influenced researchers of the world who contributed 
to the development of today's quantum information science via 
the researchers trained at the international conference [30-39] that 
I established under the direction of Ikehara. 

Finally, I add a short remark.
As Wiener pointed out, in order to create new concepts 
in the future, the researchers of information science 
must lead quantum information science.  
In other words, a mathematical research is 
expected to be carried out in consideration of real communication 
system from the perspective of information science, not quantum 
statistical physics.
We have excelent researchers who have approached from information science and mathematics in Japanese community. Thus I  appreciate  the great contributions of the Japanese researchers  listed below:\\
In the entropy theoretical approaches,Masaki Sohma [40], Masahito Hayashi [41], 
Keiji Matsumoto [42, 43] and Tomohiro Ogawa and 
Hiroshi Nagaoka [44]. 
In the signal detection approaches, 
Akio Fujiwara[45], Masashi Ban [46], Masao Osaki[47],
Kenji Nakahira[48], Kentaro Kato[49],  Tsuyoshi Usuda[50], Jun Suzuku [51].
In the coding theory approach, Mitsuru Hamada [52] .

\appendix
\subsection{Proof of Theorem 7} 
Let us consider a guessing probability of real system 
and that of the ideal case. Now we can apply the theory of multi-hypothesis 
quantum detection by Holevo [2]-Yuen [3]. 
A set of quantum states in the signal space
$H_S$ is given as $\rho_i \in H_S$, $i=1,2,3,\dots M $.
The criterion of quantum detection strategy is as follows:
\begin{eqnarray}
&&P_e=\min\limits_{\bf{\Pi}}
(1-\sum_{i=1}^M p(i) Tr {\Pi}_i{\bf{\rho} }_i )\\
&&P_d=1-P_e
\end{eqnarray}
where $P_e$ and $P_d$ are average error probability and detection probability.
Here, I employ Portman's method [53] as a basis.
First, we apply the above to find the detection probability in QKD system with 
quantum composite system.
Let us consider two different cases such as detection probability 
of real one, and that of the ideal case, and compare both detection probabilities. 
A set of density operators for the real one is given by $\{\rho_{KE}^k\}$ 
and a set of the ideal one is given by $\rho_{K}\otimes \rho_{E}$,
where $\{k \in {\cal M}\}$. 
Each detection probability in two cases is deduced by using the formula of 
Eq(33) or Eq(34). But, here let us assume that 
$\Lambda^+ =(\sum_k|k><k| \otimes \Pi_k^{opt})$ 
as sub-optimum POVM in composite system, and the density operator for 
the ideal one is $\rho_K\otimes \rho_E=\sum_k (1/M)|k><k|\otimes \rho_E$. 
Then the detection probability 
of real case $P_d^R$ and that of the ideal $P_d^I$ are 
\begin{eqnarray}
P_d^R&=&Tr\Lambda^+ p(k)\rho_{KE}^k\\
P_d^I&=&Tr\Lambda^+ \rho_{K}\otimes \rho_{E}=\frac{1}{M}
\end{eqnarray} 
Since the trace distance $\Delta_q$ is defined by Eq(31) as the maximum with 
respect to any POVM, the trace distance between Eq(35) and Eq(36) satisfies
\begin{equation}
Tr\Lambda^+ (\sum_k p(k)\rho_{KE}^k-\rho_{K}\otimes \rho_{E}) \le \Delta_q
\end{equation}
Hence we have the following one from Eqs(33-36) as the upper bound 
of detection probability for the real system: 
\begin{equation}
P_d^R=Tr\Lambda^+ p(k)\rho_{KE}^k \le \frac{1}{M} +\Delta_q
\end{equation}
Then, the average guessing probability is given by 
\begin{eqnarray}
P(K)_{guess}&=&\max \sum_{i=1}^M p(y_i)p(x_i|y_i) \nonumber \\
&=&\max \sum_{i=1}^M p(i) p(y_i|x_i) \nonumber \\
&=& P_d
\end{eqnarray}
The lower bound of the detection probability is 
simply $1/M$ as the pure guessing in the signal detection theory.
It can be given by $\Delta_q=0$ which means that the real case is 
equal to the ideal case [22,26].
That is, there is no correlation between key sequence $K$ and 
observation data $E$ of Eve. So one can denote associated with 
 Shannon theory in the perfect case as follows:
\begin{equation}
P(K|E) =P(K), \quad or \quad H(K|E)=H(K)
\end{equation}


\begin{thebibliography}{99}
\bibitem{ref001}
C.~W.~Helstrom, 
{\sl Quantum Detection and Estimation Theory}
(Academic Press, 1976).
\bibitem{ref002}
A.~S.~Holevo,
``Statistical decision theory for quantum systems,"
{\it J. Multivar. Anal.},
vol. 3, no. 4, pp. 337-394, Dec. 1973.
\bibitem{ref003}
H.~P.~Yuen, R.~S.~Kennedy, and M.~Lax,
``Optimum testing of multiple hypotheses in quantum detection theory," 
{\it IEEE Trans. Inf. Theory}, 
vol. 21, no. 2, pp. 125-134, Mar. 1975.
\bibitem{ref004}
H.~P.~Yuen and J.~H.~Shapiro, 
``Quantum statistics of homodyne and heterodyne detection,"
In: L.~Mandel and E.~Wolf E. (eds), 
{\sl Coherence and Quantum Optics IV} 
 (Plenum Press/Springer, Boston), 
 pp. 719-727, 1978.
\bibitem{ref005}
O.~Hirota,
``Generalized quantum measurement theory and its application for 
quantum communication theory --- optical communications by two-photon laser ---," 
{\it Trans. IECE of Japan},
vol. 60-A, no. 8, pp. 701-708, Aug. 1977; in Japanese. 
\bibitem{ref006}
R.~Jozsa, D.~Robb, and W.~K.~Wootters,
``Lower bound for accessible information in quantum mechanics,"
{\it Phys. Rev. A}, 
vol. 49, no. 2, pp. 668-677, Feb. 1994.
\bibitem{ref007}
M.~Ban, M.~Osaki, and O.~Hirota, 
``Upper bound of accessible information and lower bound of Bayes cost 
in quantum signal detection process,"
{\it Phys. Rev. A},  
vol. 54, no. 4, pp. 2718-2727, Oct. 1996;
and also 
M.~Ban, K.~Yamazaki, and O.~Hirota,
``Accessible information in combined and sequential quantum measurement 
on binary state signal," 
{\it Phys. Rev. A},
vol. 55, no. 1, pp. 22-26, Jan. 1997.
\bibitem{ref008}
O.~Hirota, 
{\sl Optical Communication Theory --- Basis of Quantum Theory ---}
(Morikita Publishing, Tokyo, 1985); in Japanese. 
\bibitem{ref009}
A.~S.~Holevo, 
``The capacity of the quantum channel with general signal states," 
{\it IEEE Trans. Inf. Theory}, 
vol. 44, no. 1, pp. 269-273, Jan. 1998.
\bibitem{ref010}
B.~Schumacher and M.~D.~Westmoreland, 
``Sending classical information via noisy quantum channel,"
{\it Phys. Rev. A},
vol. 56, no. 1, pp. 131-138, July 1997.
\bibitem{ref011}
A.~S.~Holevo, M.~Sohma, and O.~Hirota, 
``Capacity of quantum Gaussian channels,"
{\it Phys. Rev. A},
vol. 59, no. 3, pp. 1820-1828, Mar. 1999.
\bibitem{ref012}
A.~S.~Holevo, M.~Sohma, and O.~Hirota, 
``Error exponents for quantum channels with constrained inputs,"
{\it Rep. Math. Phys.},
vol. 46, no. 3, pp. 343-358, Dec. 2000. 
\bibitem{ref013}
S.~Kullback, 
{\sl Information Theory and Statistics} 
(Dover, 1959).
\bibitem{ref014}
H.~Love, ``The Ikehara collection: Norbert Wiener's Japan connections,"
{\it IEEE Technol. Soc. Mag.},
vol. 36, no. 2, pp. 44-49, June 2017.
\bibitem{ref015}
O.~Hirota,
``What I learned from Prof. Shikao Ikehara,"
IEICE General Conference 2019,
AK-2-9, Mar. 2019; in Japanese.
\bibitem{ref016}
H.~Umegaki, 
``Conditional expectation in operator algebra, IV (Entropy and information)," 
{\it K\={o}dai Math. Sem. Rep.},
vol. 14, no. 2, pp. 59-85, 1962.
\bibitem{ref017}
H.~Umegaki's selected papers: 
{\sl Operator Algebra and Mathematical Information Theory}
(Kaigai Publications, Tokyo, 1985).
\bibitem{ref018}
H.~Umegaki and M.~Ohya, 
{\sl Theory of quantum entropy}
(Kyoritsu Shuppan, Tokyo, 1984); in Japanese. 
\bibitem{ref019}
M.~Ohya and D.~Petz, 
{\sl Quantum Entropy and Its Use} 
(Springer, 1993).

\bibitem{ref020}
M.~M.~Wilde,
{\sl Quantum Information Theory} 
(Cambridge University Press, 2016);
and
J.~Watrous,
{\sl The Theory of Quantum Information} (Cambridge University Press, 2018).
\bibitem{ref021}
F.~Hiai, M.~Ohya, and M.~Tsukada,
``Sufficiency, KMS condition and relative entropy in von Neumann algebras,"
{\it Pac. J. Math.},
vol. 96, no.1, pp. 99-109, Sep. 1981.
\bibitem{ref022}
O.~Hirota, 
``Incompleteness and limit of quantum key distribution,"
{\it Tamagawa University Quantum ICT Research Institute Bulletin} 
(open access), 
vol. 2, no. 1, pp. 25-34, 2012;
also 
arXiv:1208.2106v2 [quant-ph], 2012.
\bibitem{ref023}
H.~P.~Yuen, 
``Key generation: Foundations and a new quantum approach,"
{\it IEEE J. Sel. Top. Quantum Electron.},
vol. 15, no. 6, pp. 1630-1645, Nov./Dec. 2009.
\bibitem{ref024}
H.~P.~Yuen, 
``Fundamental quantitative security in quantum key generation,"
{\it Phys. Rev. A}, 
vol. 82, no. 6, 062304, Dec. 2010.
\bibitem{ref025}
H.~P.~Yuen,
``Security of quantum key distribution," 
{\it IEEE Access},
vol. 4, pp. 724-749, Feb. 2016.
\bibitem{ref026}
O.~Hirota,
``A correct security evaluation of quantum key distribution," 
{\it Tamagawa University Quantum ICT Research Institute Bulletin} (open access), 
vol. 4, no. 1, pp. 1-9, 2014.
\bibitem{ref027}
G.~Cariolaro,
{\sl Quantum Communications} 
(Springer, 2015).
\bibitem{ref028}
G.~C.~Papen and R.~E.~Blahut, 
{\sl Lightwave Communications} 
(Cambridge University Press, 2019).
\bibitem{ref029}
P.~Verma, M.~El Rifai, and K.~W.~C.~Chan, 
{\sl Multi-photon Quantum Secure Communication} 
(Springer, 2019).
\\
\hrulefill

{\sl The Proceedings of Conference on Quantum Communication,
Measurement, and Computing}
\bibitem{ref030}
C.~Bendjaballa, O.~Hirota, and S.~Reynoud (Eds.), 
{\sl Quantum Aspects of Optical Communications},
(Lecture Note in Physics, 378), (Springer Verlag, 1991).
\bibitem{ref031}
V.~P.~Belavkin, O.~Hirota, and R.~L.~Hudson (Eds.),
{\sl Quantum Communications and Measurement},
(Prenum Press; Springer, 1995).
\bibitem{ref032}
O.~Hirota, A.~S.~Holevo, and C.~M.~Caves (Eds.),
{\sl Quantum Communication, Computing, and Measurement}
(Prenum Press; Springer, 1997).
\bibitem{ref033}
P.~Kumar, G.~M.~D'Ariano, and O.~Hirota (Eds.)
{\sl Quantum Communication, Computing, and Measurement 2} 
(Kluwer Academic/Prenum Press; Springer, 1999).
\bibitem{ref034}
P.~Tombesi and O.~Hirota (Eds.)
{\sl Quantum Communication, Computing, and Measurement 3}
(Kluwer Academic/Plenum Press; Springer, 2001).
\bibitem{ref035}
J.~H.~Shapiro and O.~Hirota (Eds.)
{\sl Quantum Communication, Measurement and Computing} 
 (Rinton Press, 2003).
\bibitem{ref036}
S.~M.~Barnett, E.~Anderson, J.~Jeffers, P.~\"{O}hberg, and O.~Hirota (Eds.),
{\sl Quantum Communication, Measurement and Computing},
{\it AIP conference proceedings, 734},
(American Institute of Physics Press, 2005).
\bibitem{ref037}
O.~Hirota, J.~H.~Shapiro, and M.~Sasaki (Eds.),
{\sl Quantum Communication, Measurement and Computing}
(NICT Press, 2007).
\bibitem{ref038}
A.~Lvovsky (Ed.),
{\sl Quantum Communication, Measurement and Computing},
{\it AIP conference proceedings, 1110}, 
(American Institute of Physics Press, 2009).
\bibitem{ref039}
T.~Ralph and P.~K.~Lam (Eds.),
{\sl Quantum Communication, Measurement and Computing},
{\it AIP conference proceedings, 1363}, 
(American Institute of Physics Press, 2011).
\\
\hrulefill
\bibitem{ref040}
M.~Sohma, ``Capacity of quantum Gaussian channels,"
{\it Trans. IEICE of Japan}, 
vol. J88-A, no. 8, pp. 895-902, Aug. 2005; in Japanese.\\
M.~Sohma, and O.~Hirota, ``Holevo capacity of attenuation channels 
 assisted by linear amplifiers,"
{\it Phys. Rev. A}, 
vol.76, 024303, 2007.
\bibitem{ref041}
M.~Hayashi, 
{\sl Quantum Information: An Introduction} (Springer, 2006).
\bibitem{ref042}
K.~Matsumoto,
``Reverse test and characterization of quantum relative entropy,"
arXiv:1010.1030 [quant-ph], 2010.
\bibitem{ref043}
K.~Matsumoto,
``On single-copy maximization of measured $f$-divergence 
between a given pair of quantum states,"
arXiv:1412.3676v5 [quant-ph], 2016.
\bibitem{ref044}
T.~Ogawa and H.~Nagaoka, 
``A new proof of the channel coding theorem via hypothesis testing 
in quantum information theory,"
arXiv:0208139 [quant-ph], 2002.
\bibitem{ref045}
A.~Fujiwara, 
``Quantum channel identification problem," 
{\it Phy. Rev. A},
vol. 63, 042304, 2001.

\bibitem{ref046}
M.~Ban, K.~Kurokawa, R.~Momose, and O.~Hirota,
``Optimum measurements for discrimination among symmetric quantum states 
and parameter estimation," 
{\it Int. J. Theor. Phys.},
vol. 36, no. 6, pp. 1269-1288, June 1997.
\bibitem{ref047}
M.~Osaki, M.~Ban, and O.~Hirota,
``Derivation and physicalinterpretation of the optimum detection 
for coherent state signals" ,
{\it Phys. Rev. A},
vol. 54, pp. 1691-1701, Aug. 1996.

\bibitem{ref048}
K.~Nakahira, K.~Kato, and T.~S.~Usuda,
``Generalized quantum state discrimination problems,"
{\it Phys. Rev. A},
vol. 91, no. 5, 059901, May 2015.
\bibitem{ref049}
K.~Kato, M.~Osaki, M.~Sasaki, and O.~Hirota, 
``Quantum detection and mutual information for QAM and PSK signals,"
{\it IEEE Trans. Commun.}, 
vol. 47, no. 2, pp. 248-254, Feb. 1999.
\bibitem{ref050}
T.~Usuda, I.Takumi, M.Hata, O.Hirota, 
``Minimum erroe detection of classical linear codes 
sending through quantum channel ," 
{\it Physics Letters A},
vol. 256, pp. 104-108, 1999.

\bibitem{Suzuki51}
J.Suzuki, ``Entanglement detection from channel parameter estimation problem",
Phayical Review A, vol-A-94, 042306,2016.

\bibitem{ref052}
M.~Hamada, 
``Concatinated quantum codes constructible in polynomial time: 
efficient decoding and error correction," 
{\it IEEE Trans. Information Theory},
vol. 54, no. 12, pp. 5689-5704, 2008.


\bibitem{ref053}
C.~Portmann and R.~Renner, 
``Cryptographic security of quantum key distribution,"
arXiv:1409.3525 [quant-ph], 2014.
\end{thebibliography}
\end{document}